\documentstyle[mprocl,natbib,epsfig]{article}

\begin{document}

\pagestyle{plain}

\title{
Rotating Band Pion Production Targets for
Muon Colliders and Neutrino Factories
\footnote{Presented at the ICFA/ECFA Workshop
"Neutrino Factories based on Muon
Storage Rings" ($\nu$FACT'99), Lyon, France, 5--9 July, 1999.}
}

\author{B.J. King}
\address{Brookhaven National Laboratory, Building 901A,
P.O. Box 5000, Upton, NY11973\\
email: bking@bnl.gov\\
web page: http://pubweb.bnl.gov/people/bking}

\maketitle

\abstracts{
  An update is presented on a conceptual design for a pion production
target station using a rotating cupronickel band and that was originally
proposed for use at a muon collider facility with a 4 MW pulsed
proton beam. After reviewing the salient design features and
motivations for this target, ongoing studies are described
that are attempting to benchmark the thermal stresses and
radiation damage on the target band using data from the
Fermilab antiproton source and other operating targets.
Possible parameter optimizations and alternative technologies for
the rotating band are surveyed, including discussion on the
the various proton beam parameters that might be
encountered for rotating band targets at either muon colliders
or neutrino factories. Finally, an outline is proposed for a
possible R\&D path towards capability for the actual construction
of rotating band pion production targets.}


\section{Introduction and Motivation}
\label{sec:intro}

 Conceptual design studies for muon colliders that have taken place
since the mid-1990's have motivated the design of pion production
targets that can operate and survive with megawatt-scale pulsed
proton beams. Over the past few months, the design timescale and
potential learning curve for such targets has effectively been
abbreviated by the expanded interest in neutrino factories --
muon storage rings dedicated to producing neutrino beams that will
require similarly large muon currents to muon colliders and that
have prospects for being built on a shorter timescale -- perhaps
to be ready within the next decade.

 We present an update on a previous report~\cite{KMMW}
describing a conceptual
design for a cupronickel rotating band pion production target for
muon colliders that was proposed as a relatively conservative
extrapolation from existing targets. A more detailed write-up
on this target is in progress~\cite{targetnim}. (Note that
another rotating band design based on reference~\cite{KMMW}
is presented elsewhere in these proceedings~\cite{Bennett}.)

 The rotating band design is readily
adaptable to neutrino factories and, indeed, its conservative nature makes
its development particularly well matched to the shorter timescales
for this application. Alternative, more exotic, targetry options
that were originally proposed for muon colliders -- such as pulsed
mercury jets -- may involve extensive multi-year exploratory
experimental R\&D programs~\cite{E951} that do not appear to
be particularly compatible with the shorter timescales envisaged
for neutrino factories.

  This paper is laid out as follows. General design considerations
and strategies for high power solid targets form the topic of
the following section. Sections~\ref{sec:design} through
section~\ref{sec:stress} then focus in on a review of the specific
cupronickel band design that was proposed in reference~\cite{KMMW}.
Specifically, section~\ref{sec:design} gives an overview of the
conceptual design and strawman specifications, section~\ref{sec:yield}
summarizes the results of computer simulations predicting its
pion yield performance and characterizing the heating effects
from the beam, and section~\ref{sec:stress} discusses stress
and durability issues. Section~\ref{sec:options} steps back
from the parameters in~\cite{KMMW} to examine the technology options
and parameter optimizations available to the rotating
band target concept for the spectrum of possible proton
driver scenarios at both muon colliders and neutrino
factories. The paper concludes with comments on the potential
for the rotating band target concept and with an outlook on
the R\&D program required to bring this conceptual design
to practical fruition.

\section{General Design Strategies for High Power Production Targets}
\label{sec:strategies}

 This section gives an overview of the general
design goals that were considered important when proposing
the target design of reference~\cite{KMMW},
and on the strategies employed to achieve these design goals.
The intention was to design a target station that:
\begin{enumerate}
  \item  can be designed quickly and with relatively modest resources, so it
will assuredly not hold up the overall development of
neutrino factories or muon colliders
  \item  will clearly survive any beam-induced stresses it might be
subjected to and have an acceptable lifetime
  \item  is relatively straightforward and affordable to build and
maintain (including target replacement and disposal)
  \item  has pion yields per proton and phase space densities that
are as good as, or not much inferior to, what could be achieved with
the more idealized targets that could be designed for operating with
low beam powers.
\end{enumerate}

 The first item implies that the R\&D program
for such a target can be conducted mainly through paper studies,
computer simulations and engineering computer-assisted design studies,
perhaps augmented by a modest amount of mechanical prototyping and/or
beam tests if convenient. Such an R\&D program appears plausible 
for the strawman design scenario of reference~\cite{KMMW}. Items 3 and 4
also appear to be met by this scenario, as can be judged from
the information presented in section~\ref{sec:design} of this
paper and elsewhere~\cite{KMMW,targetnim}.

 The remainder of this section addresses item 2 in the list, since
the single most difficult design constraint for rotating bands
and other solid targets for muon colliders is the requirement
of survivability in the face of instantaneous beam
energies per proton pulse of order 100 kJ and megawatt-scale beam powers,
i.e., comparable to or larger than the largest existing proton facilities.
A sound design strategy to satisfy item 2 is to choose beam and target
parameters such that the maximum material stresses and radiation
exposures do not go beyond what has been explicitly achieved in
existing targets or targetry studies. It is clear that this can
always be achieved in principle, even for the highest beam
powers under consideration, by:
\begin{enumerate}
  \item choosing a target material with appropriate mechanical
properties and moderate thermal stresses, consistent with optimizing
the pion yield. In practice this might mean choosing from materials
with medium atomic numbers in the range from titanium (${\rm Z = 22}$)
through nickel (${\rm Z=28}$), as discussed further in
sections~\ref{sec:yield} and~\ref{sec:options}, and
  \item sufficiently spreading out the spot size to cope with
higher energy beam pulses, and
  \item  rotating or otherwise moving the target to
continually expose new areas
of the target to each beam pulse. This limits both the instantaneous
local thermal stresses and the lifetime radiation exposure of the
target material.
\end{enumerate}

 Concerning the final item of strategy, the considerable potential
for designs that continually expose the beam to new target material
is amply illustrated by the tungsten target design for the proposed
Accelerator Production of Tritium (APT) project~\cite{APT} at
Los Alamos National Laboratory. A 1.7 GeV, 100 mA beam 
is continually rastered across a 19 x 190 cm area of the APT
tungsten target to limit the local heating and thermal shock stresses.
Impressively, the projected 170 MW beam power onto the APT target
is approximately two orders of magnitude larger than the applications
considered in this paper.

 As a detailed difference from the approach for the APT,
the band target uses the more traditional solution for high power
targets of moving the target material rather than the beam spot.
The idea of rotating or scanning high-power targets has already
found successful application in several existing facilities including,
for example, the trolled tungsten-rhenium
target for the SLC positron source~\cite{SLC} and the rotating nickel
target at the BNL ``g-minus-2'' experiment~\cite{gminus2}.
Target rotation can limit temperature rises and the consequent
thermal stresses on timescales of a second or
less and, in the longer term, it spreads out the radiation load
over much more material.
The Fermilab antiproton source target~\cite{antiproton} is rotated
more slowly than the preceding examples in order to spread out
the radiation dose over the circumference of the target disk,
as is discussed further in section~\ref{sec:stress}.

\section{A Conceptual Design and Straw-man Specifications for a
Cupronickel Rotating Band Target}
\label{sec:design}

\begin{figure}[t!] 
\centering
\epsfig{file=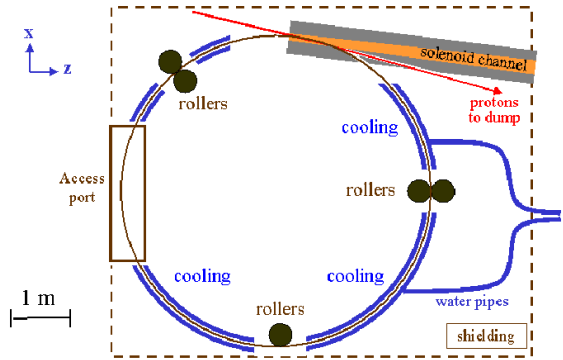, angle=270, width=3.0in}
\caption{A conceptual illustration of the targetry setup
proposed in reference~\cite{KMMW}.}
\label{layout}
\end{figure}

\begin{figure}[t!] 
\centering
\epsfig{file=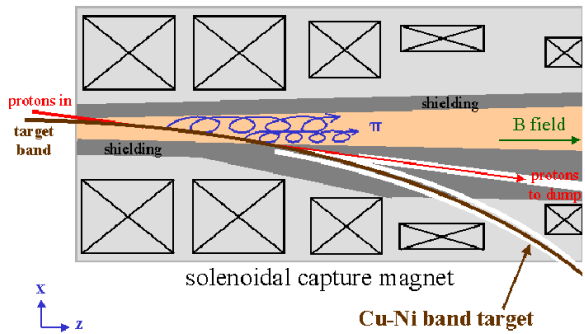, angle=270, width=3.0in}
\caption{A conceptual illustration~\cite{KMMW}
of the target layout around the pion production region.}
\label{closeup}
\end{figure}

\begin{figure}[t!] 
\centering
\epsfig{file=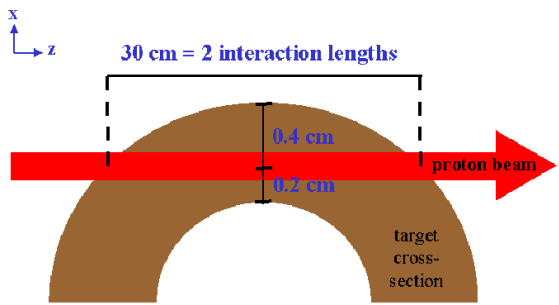, angle=270, width=3.0in}
\caption{The trajectory of the proton beam into the cupronickel
target band~\cite{KMMW}.
The aspect ratio is very distorted -- only a small chord of the target
circumference is shown.}
\label{band}
\end{figure}

\begin{table}[htb!]
\caption{Strawman parameters for the cupronickel target band,
taken from reference~\cite{KMMW}.}
\begin{tabular}{|r|c|}
\hline
target band radius (R)        &  2.5 m    \\
band thickness (t)            &  0.6 cm  \\
band width (w)                &  6 cm     \\
beam path length in band (L)  &  30 cm    \\
proton interaction lengths ($\lambda$)    &  2  \\
band tilt angle ($\alpha$)    &  150 mrad     \\
band rotation velocity (v)    &  3 m/s  \\ 
\hline
\end{tabular}
\label{tab:band}
\end{table}

\begin{table}[htb!]
\caption{The proton beam parameters that correspond to the
strawman target parameters of table~\ref{tab:band}.
Except for the beam spot size, which is specific to the
target design, these parameters have been taken from
reference~\cite{SR} and are appropriate
for the muon collider specifications in that reference.}
\begin{tabular}{|l|l|}
\hline
parameter                     &  value \\
\hline
beam energy                   &  16 GeV \\
protons per pulse (ppp)       &  $1.0 \times 10^{14}$ \\
pulse energy                  &  256 kJ \\
pulse duration                &  $~$instantaneous ($<<\:1\:\mu$sec) \\
pulse repetition rate         &  15 Hz \\
beam power                    &  3.84 MW \\
gaussian spot size            &  $\sigma_x=1.5$ mm, $\sigma_y=10$ mm \\
\hline
\end{tabular}
\label{tab:beam}
\end{table}

  The discussion presented in this section and the two that follow
are specific to the cupronickel rotating band target design
proposed in reference~\cite{KMMW}, beginning with the overview
of that conceptual design given in this section.

 Figure~\ref{layout} gives a schematic overview of the target concept
presented in reference~\cite{KMMW}
and figure~\ref{closeup} zooms in on the production region.
It should be emphasized that such details as the rollers and cooling
setup are shown only schematically and that no concerted effort has
been put into their design or layout. 
The cupronickel target band is enclosed in a 20 Tesla
solenoidal magnetic pion capture magnet whose general design
has previously been studied~\cite{SR}
by the Muon Collider Collaboration.
The pion secondaries spiral along the solenoidal
magnetic channel before decaying into the muon bunches needed for cooling,
acceleration and injection into the collider ring.
 The radius of the solenoidal
channel, 7.5 cm, and the magnetic field strength, 20 Tesla, are those
commonly assumed for recent muon collider studies~\cite{SR}. The design
modification specific to this particular geometry concerns the provision
of entry and exit ports for the target band. Example coil
geometries~\cite{Weggel_coils} show that these ports can be
provided with little modification to the design of the solenoidal
channel.

Table~\ref{tab:band} gives some relevant parameters for the
cupronickel band and figure~\ref{band} illustrates
the trajectory of the proton beam into the target band.
The band parameters correspond to the proton beam parameters
for a muon collider that are given in table~\ref{tab:beam}.
Different parameter values might be appropriate for alternative
beam scenarios at muon colliders or for neutrino factories.
The geometry of the band is chosen to approximately maximize the
pion yield. The general requirements are that the proton pathlength through
the target material should be approximately 2 nuclear interaction lengths
and that the band should be thin enough to allow most of the pions
to escape the target. To optimize the pion yield~\cite{KMMW}, the
trajectory of the beam through a chord
of the target band is at a tilt angle of 150 milliradians to the
axis of the solenoid.

  The cupronickel band is
guided and powered by several sets of rollers that can be connected by
driveshafts to remotely housed motors outside any radiation shielding.
This scenario has been taken from the Zenzimmer mills that are used for
pressing metal sheets and has the attraction of being mechanically very
simple in the high radiation area surrounding the production region.
Bennett~\cite{Bennett} even suggests the total elimination of moving
parts other than the target band by using electromagnetic guidance
and rotation of the target band by linear motors.
Procedures for installation and extraction of the target band are
proposed in reference~\cite{targetnim}.

  The rotation rate of 3 m/s given in table~\ref{tab:band}
corresponds, for the 15 Hz beam frequency, to a target advance per
pulse by 1/3 of the chord spanned by the proton beam. The three
overlapping proton
pulses in any part of the band imply~\cite{KMMW} a maximum total temperature
rise approximately double the instantaneous rise from each
individual pulse. Temperature rises and stresses will be further
discussed in section~\ref{sec:stress}.

  A competing concern to the target heating stress that limits the
acceptable rotation rate of the target is the eddy currents
induced by the rapidly rotating band in a strong magnetic field. The 
eddy current power is proportional to the conductivity and to the
square of the band velocity and a very approximate analytic
calculation~\cite{targetnim}
predicts that the power dissipated will be of order several kW.
This is more than an order of magnitude below the beam
heating and so is clearly a manageable heat source. The power will have
to be supplied by the electric motor driving the rotation of the target
and the requirements on the drive mechanism are within typical operating
parameters for Zenzimmer steel mills.

  Cupronickel alloys are preferred over both copper and nickel for the
particular targetry application considered in this paper because their
lower electrical conductivity
will reduce the eddy currents from rotation through the magnetic
field of the solenoidal capture magnet. Cupronickel alloys
such as, for example, alloy 715 produced by Olin
Brass~\cite{olin} have essentially the same density and interaction
lengths as copper and nickel and have very similar mechanical properties.
However, Olin alloy 715 has an electrical conductivity at 20 degrees
centigrade of only 2.6 MS/m, compared to 58 MS/m for copper and 14 MS/m
for nickel.

  The hot portion of the band is rapidly carried away from the production
region into a water cooling channel, as is the case for the nickel
production target at the BNL g-2 experiment~\cite{gminus2}. Although both the
peak temperature and power are much larger than in the g-2 experiment,
the surface area of the band has been chosen large enough to give
heat transfer rates of approximately
30 ${\rm W/cm^2}$~\cite{KMMW} that are comfortable
even in the presence of a steam boundary layer. It has been
suggested~\cite{Bennett} that the target and capture channel should
be in a helium atmosphere to allow the easy distillation of water
vapor and any other impurities. This will have negligible
effect~\cite{targetnim} on pion production due to helium's low density
and low atomic number.

\section{Pion Yield Predictions and Thermal Parameters of the Cupronickel
Band Target}
\label{sec:yield}

\begin{figure}[hbt]
\centering
\epsfig{file=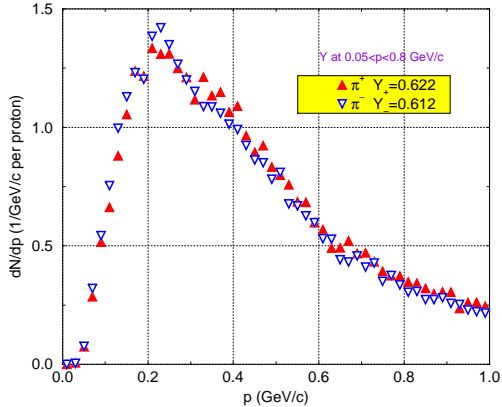, width=3.0in}
\caption{Pion momentum spectra~\cite{KMMW} at the plane 90 cm downstream from
the central intersection of the beam with the target, determined
for the beam and target parameters of tables~\ref{tab:beam}
and~\ref{tab:band}, respectively.}
\label{ppion}
\end{figure}

\begin{table}[htb!]
\caption{A summary of the MARS Monte Carlo results and derivative
predictions for pion yields and thermal parameters that were presented
in reference~\cite{KMMW}. The simulation
used the target band and proton beam parameters of
tables~\ref{tab:band} and~\ref{tab:beam}, corresponding to a 3.84 MW
proton beam power.}
\begin{tabular}{|r|c|}
\hline
parameter                     &   CuNi \\
\hline
$\pi^+$ ($\pi^-$) yield/proton  &   0.622 (0.612)  \\
peak energy deposition/pulse    &   69 J/g         \\
peak inst. temperature rise     &   $151{\rm^o}$C       \\
peak total temperature rise     &   $\sim 300{\rm^o}$C  \\
total power deposition in target band & 0.324 MW   \\
average cooling rate           &  31 W.cm$^{-2}$   \\
\hline
\end{tabular}
\label{tab:MARS}
\end{table}

  Detailed MARS~\cite{MARS} tracking and showering
Monte Carlo simulations were performed~\cite{KMMW} to obtain the pion
yields per proton for the beam and target parameters
of~\cite{KMMW} that are reproduced in tables~\ref{tab:band}
and~\ref{tab:beam}. The graphical results in figure~\ref{ppion}
correspond to yields of $Y_+$ = 0.622 and $Y_-$ = 0.612
positive and negative pions per proton for the momentum range
0.05$<$p$<$0.80~GeV/c. The peak energy deposition density was found
to be 68.6~J/g per pulse, corresponding to a temperature rise of
$\Delta T$=151$^{\circ}$C and a total power dissipation in the
target of 0.324~MW. These predictions are summarized in
table~\ref{tab:MARS}.

  These pion yields and the predicted phase space densities are
almost identical to the best predicted yields for the
exotic liquid mercury jet targets that are also under
consideration~\cite{E951} for muon colliders and neutrino
factories. Optimization studies~\cite{Snowmass, SR} for both
mercury jet and band targets further suggest that such yields
are rather close to the optimum that could be obtained even
for a low power proton beam, thus satisfying the fourth of
the design goals stated in section~\ref{sec:strategies}.

\section{Stress and Durability Issues for the Cupronickel Band}
\label{sec:stress}

  As an application of the design strategy presented in
section~\ref{sec:strategies}, this section examines the
possibilities for evaluating the survivability of the
cupronickel band target through benchmarking to existing
targets. The Fermilab antiproton source target~\cite{antiproton}
appears to be one of the most suitable targets for benchmarking
the band, so we begin by summarizing its design and operating
parameters.

  The Fermilab antiproton target consists of~\cite{antiprotondrawing}
a vertical stack of 3 nickel target disks plus one copper target disk,
each approximately 1 cm thick and 4.7 cm in radius, and interspersed
with copper cooling disks. The target is cooled by forcing air
up the vertical axis and through channels in the copper cooling
disks. The stack of disks is enclosed in a titanium
can that would contain target the material in case of failure,
but the can is not in contact with the disks and so is
irrelevant for considerations of mechanical survivability.
A 120 GeV proton beam passes through a chord of the selected
target disk, with an intensity~\cite{antiproton} of
1.6 to 2.1$\times$10$^{12}$ protons per 1.6\,$\mu$sec pulse,
which is incident every 2.4\,seconds.

 The energy per pulse of the antiproton source, up to 40 kJ,
is almost an order of magnitude below the 256 kJ muon collider
specification of table~\ref{tab:beam}.
Despite this, the round beam spot has a gaussian sigma
at entry of only 140 microns~\cite{antiproton}, and this
exposes the antiproton target
to local energy deposits and temperature rises much larger
than MARS predictions for the cupronickel band, which assume
a much larger elliptical spot of dimensions $\sigma_x=1.5$ mm,
$\sigma_y=10$ mm.
 As shown in
table~\ref{tab:band}, the cupronickel band is predicted to
sustain a maximum instantaneous energy
deposition of approximately 70 J/g per proton pulse, compared to
the 500-600 J/g maximum depositions at the the Fermilab antiproton
source. The impressive peak temperature rise in the
antiproton target is 1100${\rm^oC}$ over 1.6\,$\mu$sec,
to be compared to 150${\rm^oC}$ instantaneously and
approximately 300${\rm^oC}$ over a fraction of a second
for the band.

  While the comparison of the preceding paragraph is suggestive
that the instantaneous heat stresses on the cupronickel band might
be acceptable, there are several issues to be resolved before one
has confidence in the benchmarking comparison with the Fermilab
antiproton target. Issues include the effects of the different
target geometries and the question of how closely the
1.6\,$\mu$sec timescale for energy deposition in the antiproton
target approximates the instantaneous energy deposition in
the band.

 To further understand the shock heating stresses on the cupronickel
band,  finite element computer simulations
have been performed~\cite{Mosercomm} using ANSYS, a commercial
package that is very widely used for stress and thermal calculations.
Energy density distributions for the simulations were generated using
the MARS~\cite{MARS} particle production Monte Carlo package.
The very preliminary simulations show periodic returns to maximum stress
(i.e. ``ringing'') but with little or no amplification beyond the
initial stress. This is encouraging, and seems plausible given
that the oscillations occur in the band dimension that is much
shorter than the other two
so the geometry is quasi-one dimensional and with little potential
for focusing. Further ANSYS simulations of band target geometries are
commencing~\cite{Denshamcomm}, and these are intended to include
explicit benchmarking simulations on the Fermilab antiproton target.

 The other threat to the survival of the target band comes from
exposure to radiation. This can change the material properties of target
materials by transmuting some of the target atoms to new isotopes
or elements and by causing dislocations in the atomic lattice.
Face-centered cubic metal lattices
such as copper and nickel are known from experience to survive
radiation damage better than body-centered cubic metals like iron
or tungsten. Target damage studies at Los Alamos National Laboratory
and elsewhere~\cite{Nikolaicomm} predict a risk of failure for
copper and nickel targets after integrated doses somewhere in the
range $10^{21}$
to $10^{22}$ minimum ionizing particles per square centimeter,
which corresponds to approximately 0.3 -- 3 GJ/g of deposited
energy.

 It is straightforward to obtain a rough estimate of the
integrated doses on the cupronickel by noting that the
parameters of tables~\ref{tab:band} and~\ref{tab:beam}
correspond to doses, along the centerline of the band and for
an accelerator year of $10^7$ seconds, that accumulate to:
\begin{equation}
{\rm summed\:energy\:deposition}\: \sim
  140{\rm \:J/g\:\times \frac{3\, m/s}{2\pi\times2.5 m} \times 10^7s
           \sim 0.3\:J/g},
\label{eq:dosesum}
\end{equation}
where 140 J/g is the energy deposited per rotation and the
second term is the target rotation frequency.
Since 0.3 GJ/g is the lower limit for predictions of target failure,
the initial conclusion to be drawn is that it may well be acceptable
to replace the target band after each year's running. More detailed
studies are obviously needed to check and refine this first
simple estimate.

 To recap, this section has provided a first look at benchmarking
the cupronickel band target to the Fermilab antiproton target and
other existing data from operating targets.
The two indications from these initial comparisons are that:
\begin{enumerate}
  \item the much greater temperature rises in the operating
Fermilab target give some initial confidence in the short-term
survivability of the target but the comparison has not yet been
made rigorous
  \item the simple calculation of equation~\ref{eq:dosesum} suggests
that potentially damaging radiation doses would accumulate over
an acceptably long timescale for the cupronickel band target,
despite the 3.8 MW beam power, because the radiation dose is
spread over the entire circumference of the band rather than
being concentrated in one region.
\end{enumerate}
More detailed studies are beginning to check and clarify
these very preliminary findings.


\section{Technology Options and Parameter Optimizations}
\label{sec:options}

 The parameters of the cupronickel band target presented in
reference~\cite{KMMW} and section~\ref{sec:design} represent
no more than an educated guess at a relatively optimal
configuration for the 4 MW proton driver parameters of
table~\ref{tab:beam}. The R\&D required to verify and refine
this design is just beginning and it is expected that other
parameter values and design
refinements will likely turn out to be better suited for this
and other proton driver scenarios.

 Entry-level neutrino factories have been discussed with proton
driver powers of 1 MW or less, and this will clearly allow some
relaxation of target design parameters. At the other end of the
scale, proton drivers for neutrino factories of up to 20 MW have
also been discussed at CERN. In some neutrino factory scenarios,
the proton beam is partitioned into smaller bunches than is feasible
for muon colliders and this will generally also allow for a
relaxation of target parameters.

 Some examples of parameters that need to be optimized depending
on the specific proton driver scenario are:
\begin{enumerate}
  \item the beam spot size and the cross-sectional area of the
band. These will tend to become larger for increasing proton
pulse energy
  \item the target circumference and rotation rate: these will
both tend to increase with increasing average beam power; the
first to increase the surface area available for cooling the
target and the second to moderate the localized target heating.
\end{enumerate}
These parameter values and the those of the proton driver will
also determine the optimal technology decisions for several design
options, including:
\begin{enumerate}
  \item the cooling technology. Helium gas cooling might perhaps
be technically simpler than water cooling but the latter can
provide larger heat transfer rates per unit area. Radiative
cooling is a third possible option for refractory target
materials such as graphite~\cite{fnaltarget},
tantalum~\cite{Bennett} or tungsten
  \item the band drive and guidance. The optimal choice
might depend on the level of frictional drag from eddy
currents in the magnetic field of the capture solenoid
and this depends in turn on the target band cross section,
the target material and the target rotation velocity.
The Zenzimmer-type rollers presented in reference~\cite{KMMW}
can comfortably handle the several kilowatt frictional drag
for the cupronickel band and default parameter set. Smaller
frictional loads would instead allow for more modest
electromagnetic guidance and a linear electric drive,
as suggested in reference~\cite{Bennett}
  \item the choice of target band material, as will now
be discussed further.
\end{enumerate}

  Evaluations of several targets for pion yield and for heating
and shock stresses were performed~\cite{Snowmass, fnaltarget}
using MARS simulations. It should be noted that these comparisons
between elements  depended on target densities and on the specific
targetry scenario used, so they provide only approximate
guidance. However, the trend in heating stress was
clear, with the stresses best for low atomic number (Z) elements
and becoming rapidly worse with increasing Z. To balance this, 
pion yields were predicted to be lower for low-Z materials than
for those with medium or high atomic number. Instead of a
steady rise in yield, the yield was found to plateau somewhere
between Al (Z=13) and Cu (Z=29) (elements in between were not
investigated) and then to remain constant to within the accuracy
of the study all the way out to the high-Z elements
tungsten (Z=74) and mercury (Z=80).

  This comparative study of target elements, along with the
outstanding track record of both copper and nickel as target
materials, was part of
the basis for the choice of cupronickel as the target band
material proposed in reference~\cite{KMMW},
with nickel (Z=27) and copper (Z=28) both towards the
low-Z end of the plateau in pion yield. It would clearly
be helpful to repeat the yield study for elements between
Al and Cu to pin down the exact fall-off position
of the yield plateau, especially since some
of the elements in between and their alloys are known to have
excellent mechanical and thermal properties for targetry
applications, particularly Ti (Z=22), V (Z=23), Cr (Z=24)
and Mn (Z=25)~\cite{HERA, fusionweb}.

  While initial studies suggest the suitability of these
medium-Z elements for a 4 MW proton driver for muon colliders,
lower-Z elements such as graphite can always be considered
as options to extend the rotating band design to larger
safety margins or to even more demanding beam specifications.
The penalty to be paid is that graphite, for
example, appears to have~\cite{mokhovgraphite} only about
2/3 the pion yield of the elements on the yield plateau.
Despite this, such bands might anyway be used in the same target
station as uses medium-Z bands -- either as insurance against
problems with the medium-Z band or as an entry-level band
to be eventually replaced by one with a higher yield.

\section{Conclusions and Outlook}

  This paper has reviewed and enlarged on a previously
proposed design~\cite{KMMW} for a high power pion production
target station based around a rotating cupronickel band target.
The design scenario is mechanically rather straightforward
and can be readily extrapolated from, and benchmarked to,
existing targets. Initial computer simulations~\cite{KMMW} have
predicted relatively optimal pion yields and the initial comparisons,
made in section~\ref{sec:stress}, with the operating Fermilab
antiproton target suggest that
the cupronickel band target parameters of~\cite{KMMW} will
survive the proton beams for at least some of the neutrino
factory and muon collider scenarios that are under
discussion.  More generally, it was argued in
sections~\ref{sec:strategies} and~\ref{sec:options} that viable
rotating band target designs almost certainly exist for any of the
proton driver scenarios that have been seriously discussed
for either muon colliders or neutrino factories. The question
is one of design optimization
rather than feasibility, and whether the demands of target
survivability force any significant compromises
on the pion yield or phase space density that can be
supplied by the target.

 The evolutionary nature and relative simplicity of the target
design and concept appears to make it compatible with a
few-year R\&D program to first explore the options and
parameter space and then, if all goes well, to refine
and develop the design towards construction at a neutrino
factory or muon collider. Such an R\&D program might
develop something like this:
\begin{itemize}
  \item {\bf in 2000:} A) Paper studies to further clarify
the design options and issues, including developing 
databases of target material properties, existing targets,
past experimental targetry studies and contact information
on targetry experts and contacts, B) stress simulations and optimization
studies using ANSYS or a similar finite element analysis
package. C) beginning engineering studies on the target
layout, mechanical issues and cooling options, D) develop
a conceptual design for the beam dump, E) further
yield optimization studies on target geometry and the
band material, using MARS or similar particle production
codes,
F) further particle tracking studies to explore the integration
of the target design with the beam dump and capture and
phase rotation channels.
  \item {\bf in 2001:} More detailed design studies
for specific scenarios. Detailed assessments of, and comparisons
with, other target options.
  \item {\bf 2002 and beyond:} continuing design studies
might lead to mechanical prototyping if this is found to
be necessary and to beam tests if these are convenient and
would add significantly to the existing pool of experimental
knowledge.
  \item {\bf by approximately 2004:} ready to begin constructing
a rotating band target for a neutrino factory or muon
collider.
\end{itemize}
The R\&D issues for successive years have been spelled
out in progressively less detail but will necessarily
involve a ramp-up in manpower, beginning from perhaps
2 full-time equivalent people in 2000. Such an R\&D
program looks to be of modest extent compared to the
efforts required for the more exotic targets needed for,
e.g., neutron spallation sources. Indeed, the overall
program looks compatible with installing a rotating
band target station at a neutrino factory or muon
collider facility for potential operation as early
as 2006 or 2007.

\section{Acknowledgments}

 Studies on the cupronickel band target design have been conducted
in collaboration with R.J. Weggel, N.V. Mokhov and S.S. Moser. This
work has also benefitted from discussions and techical advice from
G. Bunce and C. Pai. The organizers and secretariat of NUFACT99 are
to be commended for a well-organized and stimulating workshop.

This work was performed under the auspices of
the U.S. Department of Energy under contract no. DE-AC02-98CH10886.

\end{document}